\documentclass[a4paper, 10pt]{article}
\usepackage[utf8]{inputenc}
\usepackage[T1]{fontenc}
\usepackage{times}
\usepackage{graphicx}
\usepackage{psfrag}
\usepackage{color}
\usepackage{amsmath}
\usepackage{amscd}
\usepackage{amssymb}
\usepackage{amsthm}
\usepackage[section]{placeins}
\usepackage[normalem]{ulem}
\usepackage{caption}
\usepackage{ragged2e}
\usepackage{subfigure}
\usepackage{nicefrac}

\DeclareCaptionJustification{reallyjustified}{\justifying}
\begin{document}

\title{Coexistence of Interacting Opinions in a Generalized Sznajd Model}
\author{Andr\'e M. Timpanaro$^1$\\ \texttt{timpa@if.usp.br}\\ Carmen P. C. Prado$^1$\\ \texttt{prado@if.usp.br} \\ $^1$ Instituto de F\'isica, Universidade de S\~ao Paulo \\  Caixa Postal 66318, 05314-970 - S\~{a}o Paulo - S\~{a}o Paulo - Brazil}
\date{\today}
\maketitle

\begin{abstract}

The Sznajd model is a sociophysics model that mimics the propagation of opinions in a closed society, where the interactions favour groups of agreeing people. It is based in the Ising and Potts ferromagnetic models and although the original model used only linear chains, it has since been adapted to general networks. This model has a very rich transient, that has been used to model several aspects of elections, but its stationary states are always consensus states.
In order to model more complex behaviours we have, in a recent work, introduced the idea of biases and prejudices to the Sznajd model, by generalizing the bounded confidence rule that is common to many continuous opinion models. In that work we have found that the mean-field version of this model (corresponding to a complete network) allows for stationary states where non-interacting opinions survive, but never for the coexistence of interacting opinions.
In the present work, we provide networks that allow for the coexistence of interacting opinions. Moreover, we show that the model does not become inactive, that is, the opinions keep changing, even in the stationary regime. We also provide results that give some insights on how this behaviour approaches the mean-field behaviour, as the networks are changed.

\end{abstract}

\section{Introduction}

In the last decade, the study of complex networks and of opinion propagation models has become increasingly common. Opinion propagation models seek to explain properties of elections, the spread of rumours and the formation of factions and consensuses in communities. These models, dubbed sociophysics models, apply approaches typical of statistical mechanics, using Monte Carlo simulations and microscopical rules.

The Sznajd model was defined by Sznajd and Sznajd-Weron under the name ``United We Stand. Divided We Fall'' as a model where the society is represented by a linear chain, and people can have one of two opposite opinions \cite{sznajd-def}. This model was quickly adapted to an arbitrary number of opinions and a general network \cite{Bernardes-2002}. The main differences between the Sznajd model and other opinion propagation models are that the interactions favour bigger groups of agreeing people and the agents can be seen to be influencing their environment, instead of the opposite (which happens in the voter model, for instance).

The transient of this model shows some interesting properties, specially when the society is modelled by a complex network, and has been used to explain some scaling properties of both proportional and majority elections \cite{Bernardes-2002}. On the other hand, the stationary states of the model are always consensus states, implying that everyone has the same opinion.

In a recent work \cite{Timpanaro-2009}, we have generalized the rules of the model, in a way that can be interpreted as the introduction of prejudices and biases. We found out that the mean-field behaviour of this generalized model allows for the coexistence of non-interacting opinions in the stationary state, and that societies modelled by Barab\'asi-Albert networks \cite{rede-BA-def} follow essentialy the mean-field behaviour.

In the present work we show that some Watts-Strogatz networks \cite{rede-WS-def}, based on square networks, allow for stationary regimes, where not only interacting opinions coexist, but the quantity of nodes holding a given opinion oscillates with time.

\section{Model Description}

We now introduce the Sznajd model as defined in \cite{Bernardes-2002}, the bounded confidence rule \cite{deffuant-def}, general confidence rules \cite{Timpanaro-2009} and a special case that can be interpreted as 2 identical parties, that have each 2 factions with different convincing powers.

\subsection{The Sznajd Model}
\label{ssec:sznajd_usual}

In the Sznajd model the society is represented by a network, where each node in the network is a person, each edge is a social connection (friendship, acquaintance, marriage, etc.) and each node $i$ is associated to an integer variable $\sigma_i$, between 1 and $M$, representing the opinion of the corresponding person.

\begin{figure}[htbp]%
    \begin{center}%
		\includegraphics[width=\textwidth]{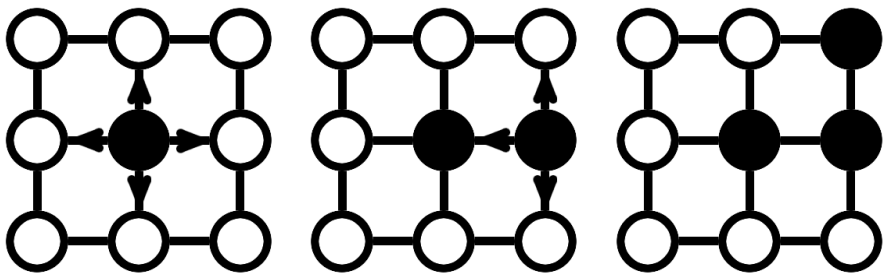}
		\caption{The Sznajd update scheme. First a node $i$ is chosen at random. Then a neighbour of $i$, $j$, is chosen. Finally, if $i$ and $j$ have the same opinion, they attempt to convince a neighbour of $j$, $k$, chosen at random.}%
		\label{fig:scheme_sznajd}%
	\end{center}%
\end{figure}

At each time step, we choose a node $i$ at random and (also at random) a neighbour $j$ of $i$. If $i$ and $j$ disagree ($\sigma_i\neq\sigma_j$), nothing happens. Otherwise, we choose at random a neighbour $k$ of $j$ and the node $k$ becomes convinced of the opinion of the pair ($\sigma_k \rightarrow \sigma_j$) with some probability $p$. In the usual model, $p$ is always 1. Figure \ref{fig:scheme_sznajd} shows a scheme of the update rule in action.

\subsection{Bounded Confidence and Confidence Rules}

The bounded confidence rule was introduced in \cite{deffuant-def} as a way to tell how close 2 continuous opinions are to each other and to model extremism. This rule says that opinion changes are not abrupt, which is accomplished by imposing that 2 agents with opinions $\sigma$ and $\sigma'$ only interact if $|\sigma - \sigma'| \leq \varepsilon$. This rule can be adapted to discrete opinion models, like the Sznajd model, by taking $\varepsilon = 1$ and changing the number of opinions instead of the interaction threshold. In the Sznajd model this means that $\sigma$ can only be convinced to an opinion $\sigma'$ iff $\sigma' = \sigma \pm 1$.

In \cite{Timpanaro-2009}, we have generalized the bounded confidence rule to what we called general confidence rules, by imposing that the probability that an opinion $\sigma$ be convinced to an opinion $\sigma'$ is given by $p_{\sigma\rightarrow\sigma'}$ (a parameter of the model). These rules can be schematized by interpreting these parameters as the elements of the adjacency matrix of a weighted directed graph, and associating this graph to the set of parameters (this graph will also be refered to as the confidence rule).

It is noteworthy that many of the previous modifications of the Sznajd Model (as well as the usual version, introduced in section \ref{ssec:sznajd_usual}) can be obtained by appropriate choices of the parameters:

\begin{itemize}
\item The usual model is the case $p_{\sigma'\rightarrow\sigma}=1$ for all $\sigma\neq\sigma'$.
\item If an opinion $\sigma$ has $p_{\sigma'\rightarrow\sigma}=0$ and $p_{\sigma\rightarrow\sigma'}=p\neq 0\,\,\forall\,\,\sigma'$, $\sigma$ behaves analogously to an undecided state.
\item Usual bounded confidence is the case $p_{\sigma'\rightarrow\sigma}=1$ if $\sigma' = \sigma \pm 1$ and $p_{\sigma'\rightarrow\sigma}=0$ otherwise.
\end{itemize}

We also introduced a phase space representation, in which the phase space variables are the proportions of sites in the network with a given opinion. We will denote these variables $\eta_{\sigma}$, that is, $\eta_{\sigma}$ is the proportion of sites in the network with opinion $\sigma$. These variables have the following constraints:

\[
\sum_{\sigma} \eta_{\sigma} = 1 \,\,\mbox{and}\,\, \eta_{\sigma} \geq 0,
\]

\noindent which imply that the phase space must be a simplex (the phase space with 2 opinions is a line segment, with 3 opinions it is a triangle, a tetraedron for 4 opinions and so on).

In the next section we investigate a family of confidence rules that can be interpreted as a dispute among 2 identical parties, that have 2 different factions each that behave differently. This family is characterized by 3 parameters, $p, q$ and $r$, and is given by:

\[
\left\{
\begin{array}{l}
p_{1\rightarrow 2} = p_{2\rightarrow 3} = p_{3\rightarrow 4} = p_{4\rightarrow 1} = p \\
p_{2\rightarrow 1} = p_{3\rightarrow 2} = p_{4\rightarrow 3} = p_{1\rightarrow 4} = q \\
p_{1\rightarrow 3} = p_{3\rightarrow 1} = p_{2\rightarrow 4} = p_{4\rightarrow 2} = r.
\end{array}
\right.
\]
 
\begin{figure}[htbp]%
    \begin{center}%
		\includegraphics[width=0.6\textwidth]{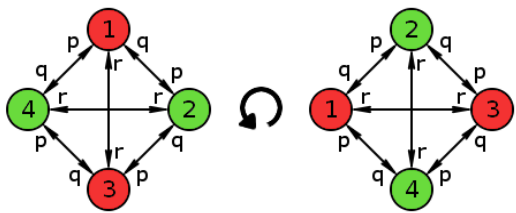}
		\caption{The 2 parties rule. Note that this graph has rotation symmetry, hence the 2 parties (green and red) are identical. The mean field results state that one of the 2 parties will always dominate the whole network and unless $r=0$, only one of the factions survive.}%
		\label{fig:rule}%
	\end{center}%
\end{figure}

Figure \ref{fig:rule} illustrates this rule, and the 2 parties $\{1, 3\}$ and $\{2, 4\}$, as well as the symmetry between them. We will denote this rule $R(p, q, r)$. The mean-field results say that if $r=0$ we have 2 attractors: $\{(\eta_1, \eta_2, \eta_3, \eta_4) | \eta_1 + \eta_3 = 1\}$ and $\{(\eta_1, \eta_2, \eta_3, \eta_4) | \eta_2 + \eta_4 = 1\}$; if $r \neq 0$ we have 4 consensus attractors ($\{(\eta_1, \eta_2, \eta_3, \eta_4) | \eta_{\sigma} = 1\}$ for $\sigma = 1,2,3,4$); and if $p=q=0$ we have a degenerate case, with 2 independent models following the original confidence rule (we will ignore this case).

\section{Simulations}

We use for these simulations a network based in the Watts-Strogatz model \cite{rede-WS-def}, but instead of rewiring a ring like network, we start from a square network with periodic boundary conditions and rewire $s.E$ edges, chosen at random ($s$ is the rewiring parameter and $E$ is the number of edges). All the networks used have, as a starting point, a square network with side $L = 316$.

For the initial conditions of the simulations, instead of drawing the opinions uniformly, we follow the approach introduced in \cite{Timpanaro-2009}. Firstly, we draw a point in the tetraedron $\{(\eta_1, \eta_2, \eta_3, \eta_4)| \eta_1 + \eta_2 + \eta_3 + \eta_4 = 1 $ and $\eta_1, \eta_2, \eta_3, \eta_4 \geq 0\}$ uniformly and then we draw the opinions 1, 2, 3 and 4 with probabilities $\eta_1, \eta_2, \eta_3$ and $\eta_4$ respectively.

\begin{figure}[htbp]%
    \begin{center}%
		\includegraphics[width=\textwidth]{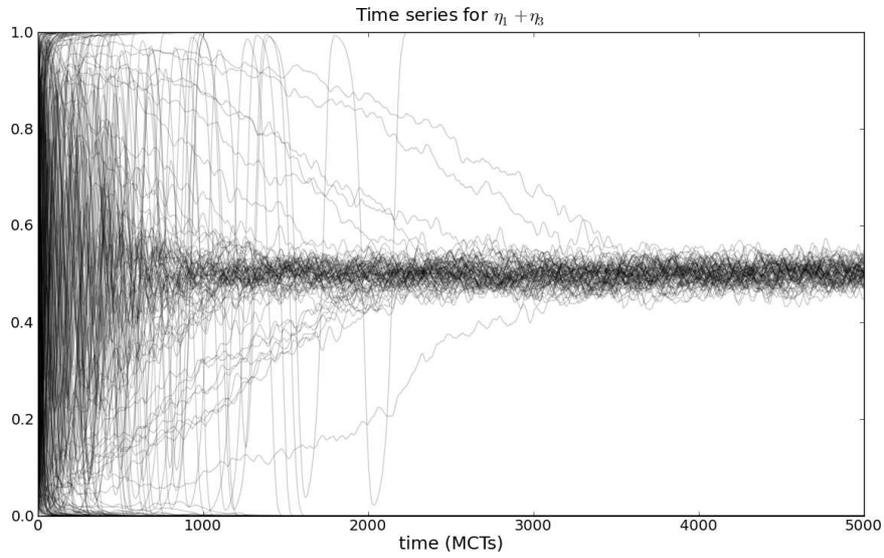}
		\caption{Time series for $\eta_1 + \eta_3$ in the case $p=1$, $q=0.1$ and $r=0$. The rewiring parameter is $s = 10^{-3}$ and the network's average path lenght is approximately 42.6.}%
		\label{fig:time_series}%
	\end{center}%
\end{figure}

We found that while some of the simulations end up in an inactive stationary regime, an oscillating (and therefore, active) stationary regime is also possible. Figure \ref{fig:time_series} shows the time series for the supporters of one of the 2 parties, using a fixed confidence rule, where this behaviour can be seen.

\begin{figure}[htbp]%
    \begin{center}%
		\includegraphics[width=\textwidth]{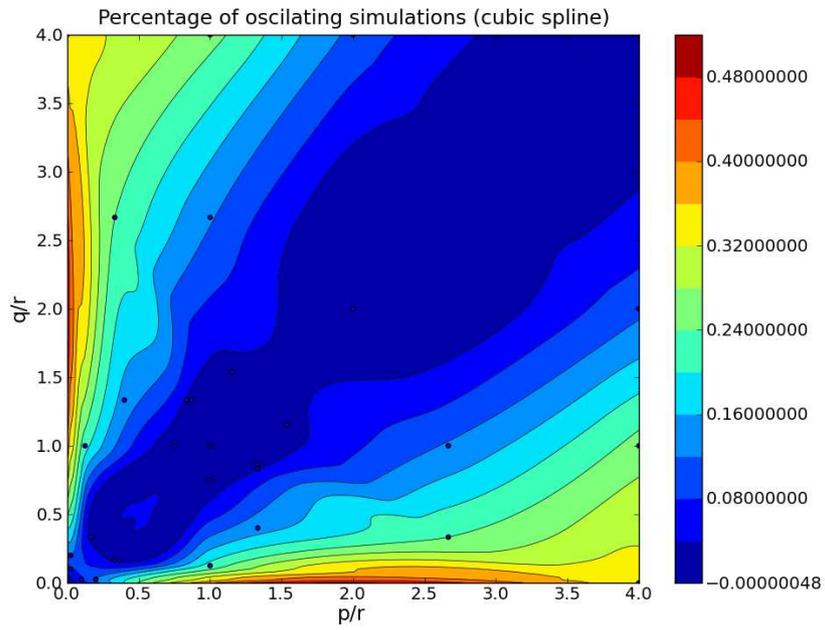}
		\caption{The percentage of simulations that end up in the oscillating stationary regime as a function of $\frac{p}{r}$ and $\frac{q}{r}$. The network is fixed, with rewiring parameter $s=3.10^{-3}$ and average path lenght approximately 29.8. The graph was obtained using cubic splines. The marks in the graph represent the actual data points.}%
		\label{fig:cubic}%
	\end{center}%
\end{figure}

The confidence rule depends on 3 parameters, however the stationary qualitative properties can be studied in only 2 dimensions, because the only difference between $R(p,q,r)$ and $R(\lambda p, \lambda q, \lambda r)$ is the time scale in which things happen. Also $R(p,q,r)$ and $R(q,p,r)$ are actually the same rule, as the corresponding graphs are isomorphic (you can go from one rule to the other by swapping opinions 2 and 4, for example). Figure \ref{fig:cubic} shows the percentage of simulations that enter the oscillating stationary regime for some values of $\frac{p}{r}$ and $\frac{q}{r}$ (the rewiring parameter $s$ was kept fixed at $s=3.10^{-3}$). It shows that these oscillations vanish when $p\simeq q$, meaning that some degree of assymetry is necessary in the confidence rule for this regime to appear.

\begin{figure}[htbp]%
    \begin{center}%
		\includegraphics[width=\textwidth]{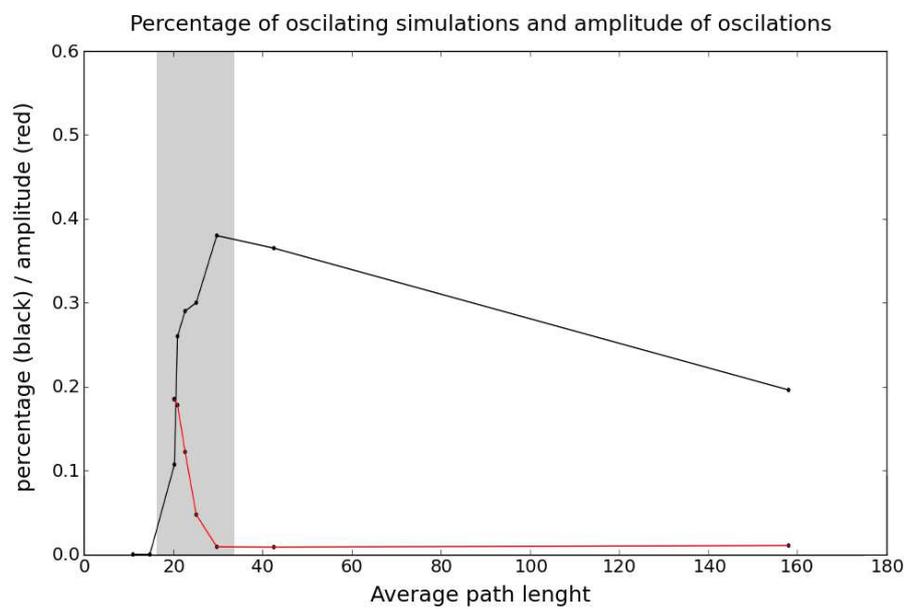}
		\caption{Amplitude of oscillations and percentage of simulations ending in the oscillating stationary regime. Note that there where no oscillating simulations found for $\ell \simeq 14.8$ and $\ell \simeq 11.0$, corresponding to rewiring parameters $s=0.03$ and $s=0.1$ respectively. The gray region is where the transition from existence to abscence of the regime is observed.}%
		\label{fig:rewiring}%
	\end{center}%
\end{figure}

We now keep the confidence rule fixed (we use $p=1$, $q=0.1$ and $r=0$) and change the rewiring parameter. Also, as the rewiring parameter grows, the average path lenght decreases and new parts of the network become connected. As we increase rewiring, the amplitude of the oscillations (which we defined as the average of the moving standard deviations over the oscillating simulations) grows, less simulations reach the oscillating stationary regime and eventually the regime ceases. Figure \ref{fig:rewiring} shows the percentage of simulations that reach an oscillating stationary regime and their amplitude as a function of the average path lenght $\ell$.

\begin{figure}[htbp]%
    \begin{center}%
		\includegraphics[width=\textwidth]{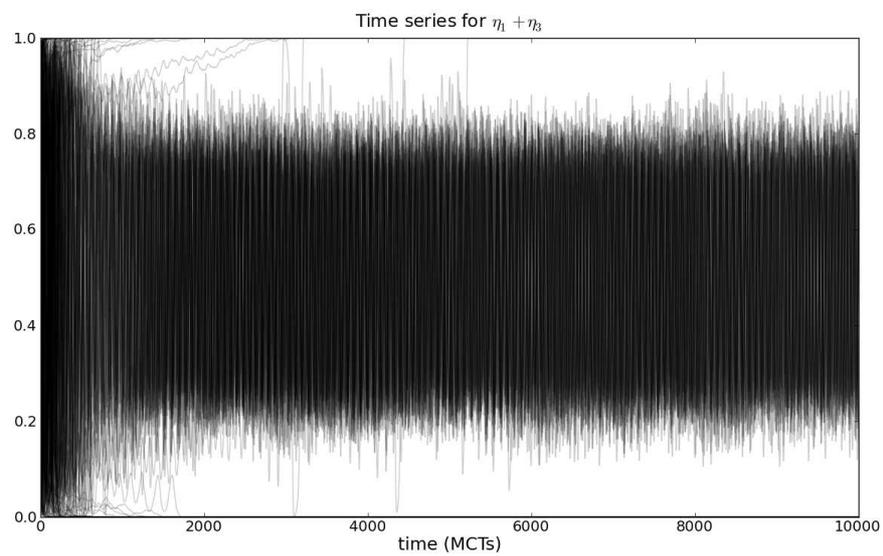}
		\caption{Time series for $\eta_1 + \eta_3$ in the case $p=1$, $q=0.1$ and $r=0$. The rewiring parameter is $s = 10^{-2}$ and the network's average path lenght is approximately 20.3.}%
		\label{fig:time_series2}%
	\end{center}%
\end{figure}

Figure \ref{fig:time_series} shows the typical behaviour of the oscillations in a case with higher average path lenght ($\ell \simeq 42.6$) and figure \ref{fig:time_series2} for a case closer to the transition to the non-oscillatory regime ($\ell \simeq 20.3$). These results can be understood from a dynamical systems perspective, using $\{(\eta_1, \eta_2, \eta_3, \eta_4)| \eta_1 + \eta_2 + \eta_3 + \eta_4 = 1 $ and $\eta_1, \eta_2, \eta_3, \eta_4 \geq 0\}$ as the phase space. We will use projections of this 3 dimensional phase space (a tetraedron) into 2 dimensions, in order to get a clearer picture of the trajectories, depicted in those time series. The phase space we are considering is a simplex, so each point in it is a convex linear combination of its vertices. Let $P_i$ be the vertices of the phase space, let $Q$ be a projection of this phase space (a linear mapping) into some vector space and let $Q_i = QP_i$. We can choose to embed our phase space in a way such that the point corresponding to $(\eta_1, \eta_2, \eta_3, \eta_4)$ is

\[
P= \sum_{\sigma} \eta_{\sigma} P_{\sigma}.
\]

\noindent So if we project the phase space, we get

\[
QP = \sum_{\sigma} \eta_{\sigma} QP_{\sigma} = \sum_{\sigma} \eta_{\sigma} Q_{\sigma}.
\]

\begin{figure}[htbp]%
    \begin{center}%
		\includegraphics[width=0.5\textwidth]{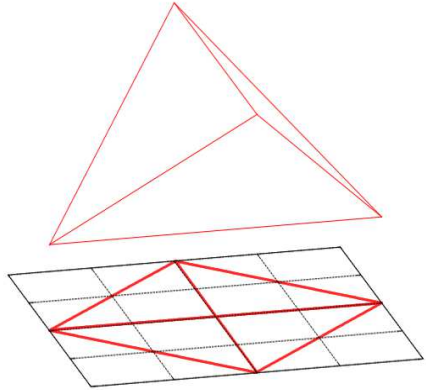}
		\caption{Scheme depicting the projection that will be used to visualize the trajectories}%
		\label{fig:proj}%
	\end{center}%
\end{figure}

\noindent So we can either choose the linear mapping $Q$ or the coordinates of the projected vertices, $Q_{\sigma}$. To build our projection of the 3 dimensional phase space in $\mathbb{R}^2$ we choose $Q_1 = (0, 1), Q_2 = (1, 0), Q_3 = (0, -1)$ and $Q_4 = (-1, 0)$, meaning that the projected phase space is a square and the time series are obtained plotting $\eta_1 - \eta_3$ against $\eta_2 - \eta_4$. Geometricaly, this corresponds to the projection depicted in figure \ref{fig:proj}.

\begin{figure}[htbp]%
    \begin{center}%
		\includegraphics[width=\textwidth]{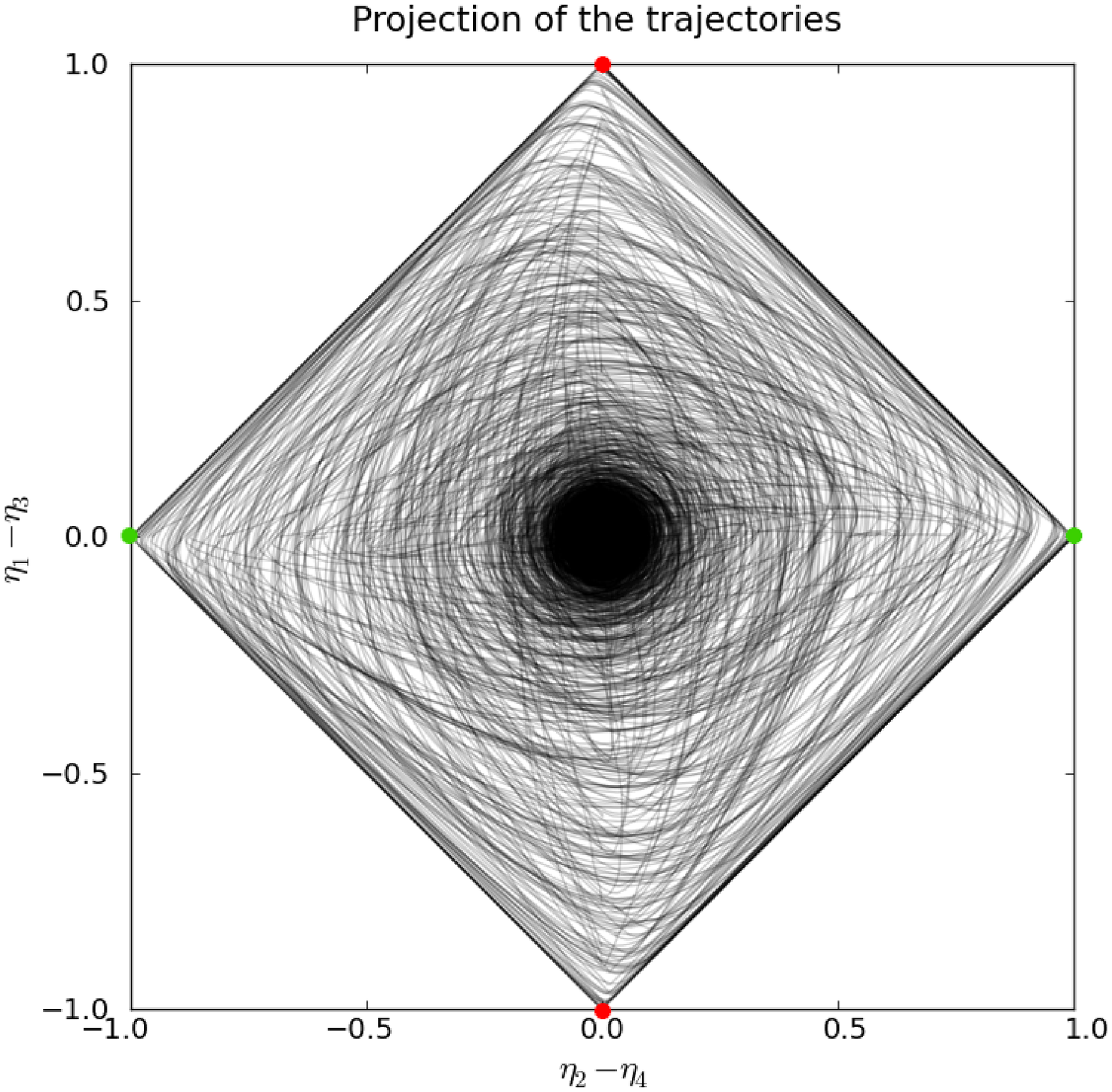}
		\caption{Projection of the trajectories in phase space for the network with $s = 10^{-3}$ and $\ell \simeq 42.6$. The projections of the mean field attractors are the line connecting the red points (coexistence of opinions 1 and 3) and the one connecting the green points (coexistence of opinions 2 and 4).}%
		\label{fig:diamond_0_001}%
	\end{center}%
\end{figure}

\begin{figure}[htbp]%
    \begin{center}%
		\includegraphics[width=\textwidth]{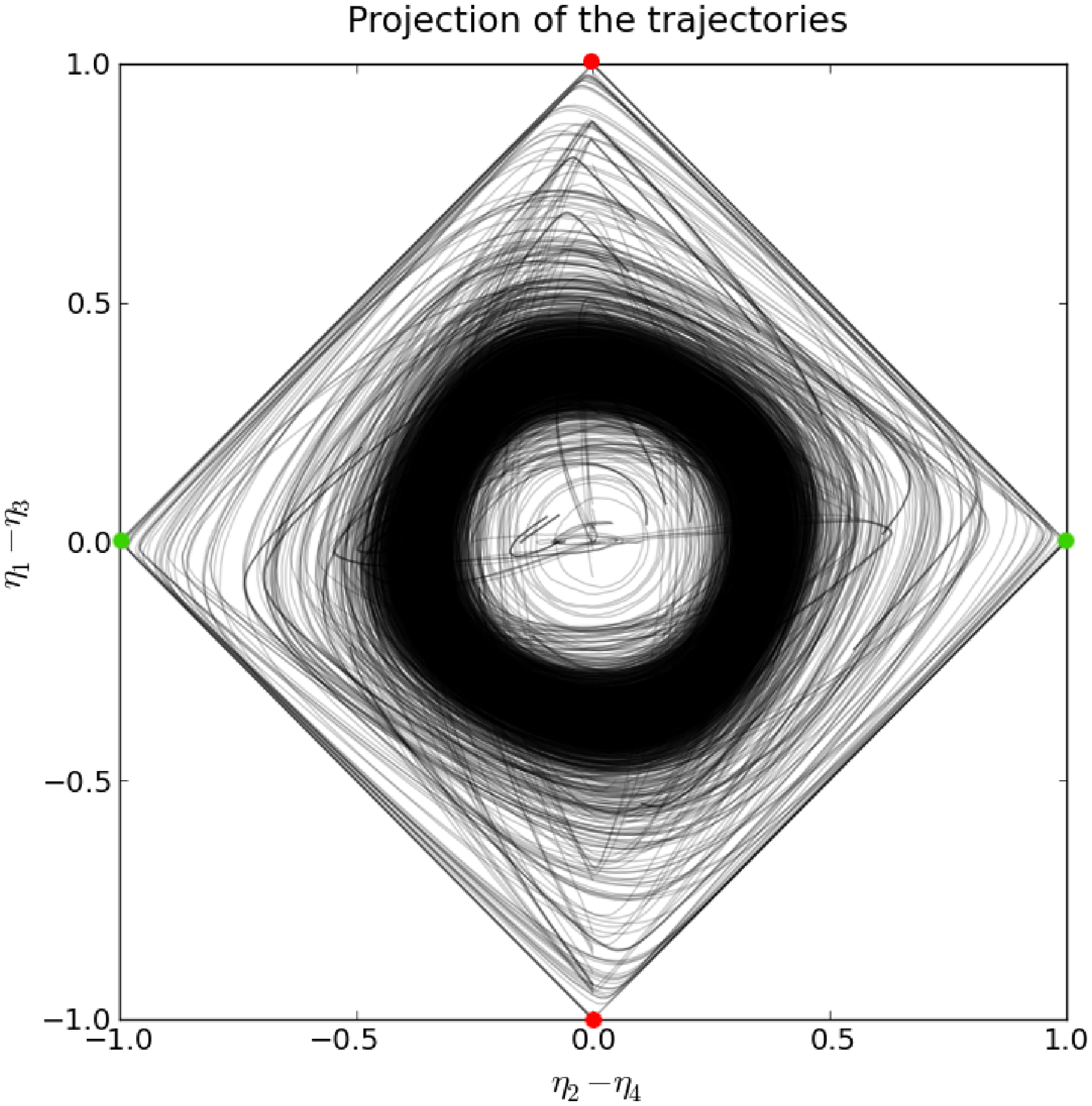}
		\caption{Projection of the trajectories in phase space for the network with $s = 3.10^{-3}$ and $\ell \simeq 29.8$. The projections of the mean field attractors are the line connecting the red points (coexistence of opinions 1 and 3) and the one connecting the green points (coexistence of opinions 2 and 4).}%
		\label{fig:diamond_0_003}%
	\end{center}%
\end{figure}

\begin{figure}[htbp]%
    \begin{center}%
		\includegraphics[width=\textwidth]{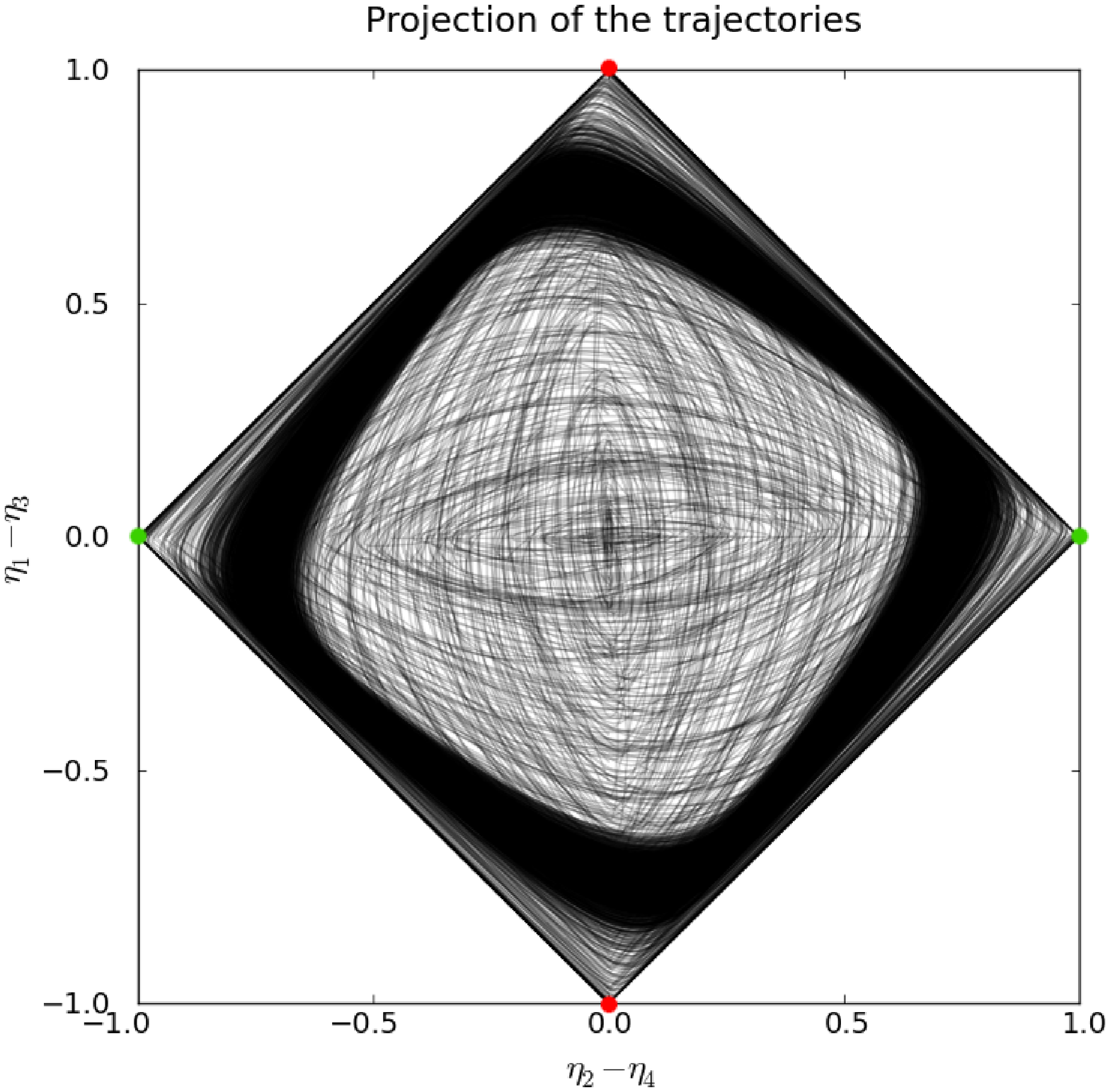}
		\caption{Projection of the trajectories in phase space for the network with $s = 10^{-2}$ and $\ell \simeq 20.3$. The projections of the mean field attractors are the line connecting the red points (coexistence of opinions 1 and 3) and the one connecting the green points (coexistence of opinions 2 and 4).}%
		\label{fig:diamond_0_01}%
	\end{center}%
\end{figure}

The oscillations in figures \ref{fig:time_series} and \ref{fig:time_series2} are due to an attractor resembling a limit cycle, whose size increases as the rewiring parameter is increased (figures \ref{fig:diamond_0_001}, \ref{fig:diamond_0_003} and \ref{fig:diamond_0_01}), but whose attraction basin becomes smaller (these 2 conditions can happen at the same time, if the basin is becoming thinner).

\begin{figure}[htbp]%
    \begin{center}%
		\includegraphics[width=\textwidth]{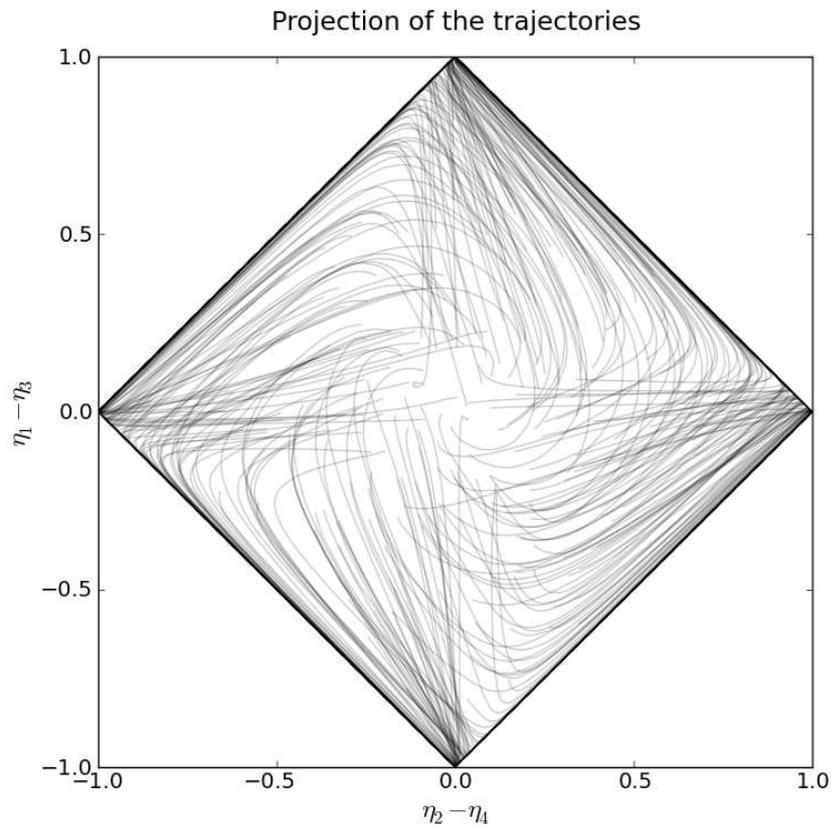}
		\caption{Projection of the trajectories in phase space for a Barab\'asi-Albert network with minimal degree 5, network size $10^5$ and $\ell \simeq 4.3$. We use for this projection $p=0.125$ and $q=r=1$. The mean field attractors are the vertices of the square.}%
		\label{fig:diamond_BA}%
	\end{center}%
\end{figure}

Finaly we present the result of these projections for a Barab\'asi-Albert network \cite{rede-BA-def}, as a comparison (figure \ref{fig:diamond_BA}). We can see trajectories connecting the attractors (the vertices of the square) which are reminiscent to the attractor responsible for the oscillating stationary regime.

\section{Conclusions}

We have shown that the Sznajd model, generalized to include prejudices and biases, can have active stationary states, with the proportion of sites holding a given opinion oscillating over time. This oscillatory regime seems to depend crucially on assymetries of the confidence rule used and on the average path lenght not being too small. The model simulated in the square lattice displays these oscillations and they remain, even if the edges are rewired (provided they are not rewired too much). The oscillations seem to be due to an attractor, resembling a limit cycle. The radius of the cycle increases as the rewiring parameter grows and the average path lenght decreases, but the volume of the basin of attraction becomes smaller, probably indicating that it is thinning. This seems to indicate that the transition between the existence and the absence of an oscillating stationary regime is due to either the limit cycle colliding with the mean-field attractors, or the basin of attraction collapsing (together with the attractor) to the frontier between the basins of the mean-field attractors (these hypothesis seem to be supported by figure \ref{fig:diamond_0_01} and to some extent by figure \ref{fig:diamond_BA}).

\vspace{1cm}
\noindent {\bf Acknowledgments}

\noindent The authors acknowledge FAPESP for financial support.

\bibliographystyle{plain}
\bibliography{andre}

\end{document}